# OPTICAL PERFORMANCE ANALYSIS OF A POROUS OPEN VOLUMETRIC AIR RECEIVER IN A SOLAR POWER TOWER

**Yu Qiu, Ming-Jia Li, Ya-Ling He[*], Ze-Dong Cheng**

Key Laboratory of Thermo-Fluid Science and Engineering of Ministry of Education, School of Energy and Power Engineering, Xi'an Jiaotong University, Xi'an, Shaanxi 710049, China

## ABSTRACT

An optical model for a solar power tower with an open volumetric air receiver (OVAR) was developed in this paper, and the optical performance and characteristics of the OVAR were studied based on the model. Firstly, the detailed distributions of the non-uniform solar flux($q_{sr}$) on the aperture and the solar source($S_{sr}$) in the OVAR were studied. The solar flux was found to be relatively uniform across the aperture for a single OVAR. However, the incident angle of the rays varies between $0°$ and $42°$, and this indicates that the parallel assumption of the rays, which is usually made in the performance analysis of OVAR, is not appropriate at this condition. Furthermore, the $S_{sr}$ in the absorber decreases from the inlet to the outlet, and the maximum source($S_{sr,max}$) of $2.414×10^8$ W·m$^{-3}$ appears at the inlet. Moreover, the $S_{sr,max}$ was found to appear at the region near the wall rather than the center of the receiver as usual for the combined effect of the non-parallel incident rays and the diffuse reflection on the wall. In addition, study on effects of the porous parameters indicates that the solid emissivity influences both the $S_{sr}$ distribution and the receiver efficiency significantly. However, the pore diameter and the porosity influence $S_{sr}$ distribution importantly, but have negligible effect on the efficiency. Finally, the optical efficiency of 86.70%, reflection loss of 13.20% and transmission loss of 0.10% were found to be achieved by the OVAR under the design condition.

## 1. INTRODUCTION

Solar power tower (SPT) using the open volumetric air receiver (OVAR) as the solar-thermal conversion module is considered as a promising technology for solar energy utilization. Because the porous absorber of the OVAR was reported to be burned down for the local high solar flux or unstable flow, studies on the solar radiation transfer in the SPT and the solar energy density distribution in the OVAR can offer help to the safe operation and accurate performance prediction of the plant.

Many studies have focused on simulating the solar radiation transfer in the heliostat field, and some codes have been developed, such as UHC, DELSOL, HFLCAL, MIRVAL, HFLD and SOLTRACE[1]. For the radiation transfer in the OVAR, some studies have also been conducted, where the solar radiation absorption on the porous absorber inlet is usually assumed to a surface phenomenon[2]. Several studies also assumed that the incident solar radiation is parallel beam and calculated the radiation transfer in the OVAR by modified P1 model[3, 4], where the effects of the solar radiation's direction distribution were ignored. Only a few studies have directly studied an OVAR in a dish collector without the above assumptions[5], but this work cannot be applied in SPT directly. It is the current status that no studies about the solar radiation transfer simulation from the heliostat field to the OVAR have been reported.

*Corresponding Author: yalinghe@xjtu.edu.cn

To provide better studies on the optical performance of the SPT with an OVAR, this work focuses on developing a comprehensive optical model using Monte Carlo Ray Tracing (MCRT)[6] method. Based on the model, the solar heat source distribution in the porous absorber, and the solar radiation transfer and absorption characteristics in the OVAR were simulated and discussed.

## 2. PHYSICAL MODEL

The DHAN heliostat field located at 40.4°N, 115.9°E in Beijing is taken as the concentrating module, where 100 heliostats are installed as shown in Fig. 1. A square OVAR is taken as the energy conversion module and is shown in Fig. 2, where the optical processes are also illustrated. The detailed parameters are given in Table 1. The optical errors of the heliostat in PS10[7] are applied in present model due to lack of DAHAN's data. Fig. 3 shows the details of the solar radiation transfer processes from the field to the OVAR.

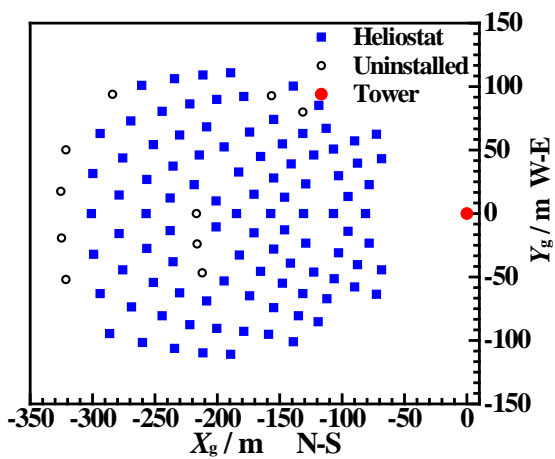
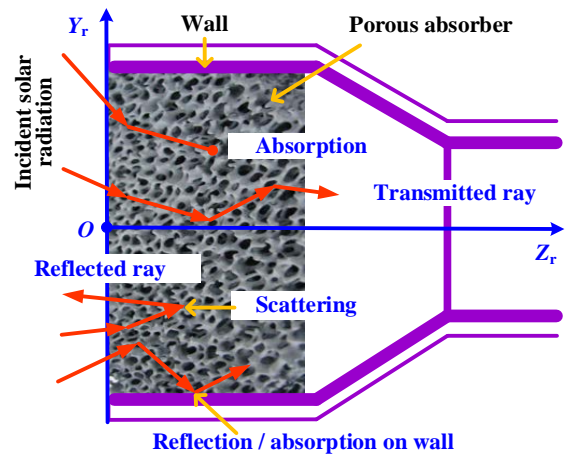

**Fig. 1** Heliostat field of DAHAN plant [8].  **Fig. 2** Sketch of the OVAR showing the radiation transfer.

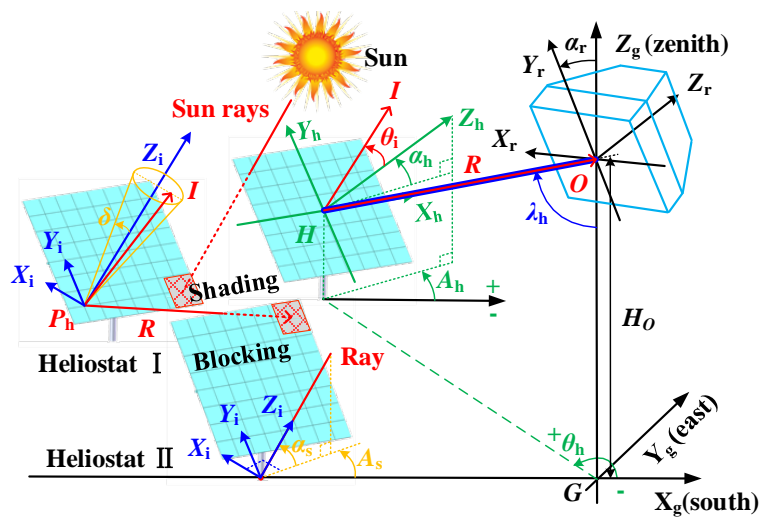

**Fig. 3** Sketch of the SPT with an OVAR showing the solar radiation transfer and coordinate systems.

Table 1 Parameters and assumptions of the model[7, 8].

| Parameters | Dim. | Parameters | Dim. |
|---|---|---|---|
| Heliostat number | 100 | Absorber thickness | 0.05 m |
| Heliostat shape | Spherical | Heliostat reflectivity | 0.90 |
| Heliostat width | 10 m | Heliostat cleanliness | 0.97 |
| Heliostat height | 10 m | Altitude tracking error $\sigma_{te,1}=\sigma_{te}$ | 0.46 mrad |
| Heliostat center height | 6.6 m | Azimuth tracking error $\sigma_{te,2}=\sigma_{te}$ | 0.46 mrad |
| Tower height | 118 m | Heliostat slope error $\sigma_{se}$ | 1.3 mrad |
| Tower radius | 10 m | SiC absorber emissivity $\varepsilon_p$ | 0.92 |
| Receiver Height $H_O$ | 78 m | Wall emissivity $\varepsilon_w$ | 0.30 |
| Receiver altitude | 25° | Porosity $\Phi$ | 0.90 |
| Absorber height | 0.14 m | Pore diameter $D_p$ | 2.0 mm |
| Absorber width | 0.14 m | | |

## 3. OPTICAL MODEL

For developing an optical model to simulate the solar radiation transfer in the system, several right-handed coordinate systems are established in Fig. 3, which are the receiver system ($X_rY_rZ_r$), the incident-normal system ($X_iY_iZ_i$), the heliostat system ($X_hY_hZ_h$), and the ground system ($X_gY_gZ_g$). For $X_rY_rZ_r$, the center of the receiver aperture $O$ is the origin. $X_r$ points to the east, and $Y_r$ points upwards. $Z_r$ is normal to $X_rY_r$ plane. For $X_iY_iZ_i$, the point hit by the ray is the origin, and $Z_i$ points towards the sun. $X_i$ is horizontal and perpendicular to $Z_i$, and $Y_i$ is normal to $X_iZ_i$ plane and points upwards. For $X_hY_hZ_h$, the center of the heliostat $H$ is the origin. $X_h$ is horizontal, and $Y_h$ is perpendicular to the tangent plane at $H$ and points upwards. $Z_h$ is normal to $X_hY_h$ plane. For $X_gY_gZ_g$, the base of the tower $G$ is the origin, and $X_g$, $Y_g$, and $Z_g$ point to the south, the east, and the zenith, respectively. The transformation matrixes among these systems can be found in our previous work[9].

Simulation of the solar radiation transfer in the heliostat field is as follows. The heliostat will track the sun when the sun shines on the field, where all the heliostats aim at the OVAR's aperture center $O$. The photons are randomly initialized on the heliostats, where the sun shape effect which induces a non-parallel angle of 4.65 mrad for the solar rays is considered[10], and the incident vector is expressed as $I_i$ in $X_iY_iZ_i$. The Direct Normal Irradiance(*DNI*) of the solar radiation is calculated by a clear sky model[11]. The reflected vector $R_h$ on the initialized location $P_h$ in $X_hY_hZ_h$ is calculated with Fresnel's law by transforming $I_i$ to $X_hY_hZ_h$, where the total equivalent slope error is calculated by $\sqrt{\sigma_{se}^2 + \sigma_{te,1}^2 + \sigma_{te,2}^2}$ [9]. The atmospheric attenuation is computed as a function of the distance between $O$ and $H$ for each heliostat[12]. The shading and blocking are also calculated, and the blocking is illustrated here to decide whether a ray is blocked or not. First, $P_h$ and $R_h$ on heliostat I are transformed to $X_hY_hZ_h$(II) as shown in Fig. 3. Then, the intersection of the ray and heliostat II's surface is calculated. Finally, if the intersection is within heliostat II, the ray is blocked. Further information about the modeling of the solar radiation transfer in the heliostat field can be found in Ref. [9].

Simulation of the solar radiation transfer in the OVAR is conducted in $X_rY_rZ_r$ as follows. When a ray hits the receiver aperture (inlet), firstly, $P_h$ and $R_h$ will be transferred to $X_rY_rZ_r$ and expressed as $P_{h,r}$ and $I_r$. Then the intersection $P_{a,r}$ in $X_rY_rZ_r$ will be calculated. The porous absorber is treated as an isotropic semitransparent media, and the solid phase is assumed to be gray and diffuse for solar radiation. The absorption coefficient ($\beta_a$), the scattering coefficient ($\beta_s$), and the extinction coefficient ($\beta_e$) of the absorber are defined in Eq.(1)[13]. Because the solar rays just transfer in the air phase of the absorber, so the refractive index (*n*) is assumed to be equal to that of air.

$$\begin{aligned} \beta_a &= 1.5\varepsilon_p(1-\phi)/D_p \\ \beta_s &= 1.5(2-\varepsilon_p)(1-\phi)/D_p \\ \beta_e &= \beta_a + \beta_s = 3(1-\phi)/D_p \end{aligned} \quad (1)$$

where $\varepsilon_p$ is the emissivity of the solid phase; $\Phi$ is the porosity; $D_p$ is the pore diameter.

When a solar photon(ray) hits the absorber aperture, the photon will interact with the porous solid phase directly. The optical interactions including scattering and absorption between the photon and the absorber are simulated in the following way. When a photon reaches an interaction site, a fraction of the photon's energy ($\Delta e_p$) will be absorbed at this site by Eq.(2) and counted in the local grid. When a photon is scattered, the travel distance (*d*) and unit directional vector ($\mathbf{R}_r$) of the scattered photon will be calculated by Eq.(3) and Eq.(4)[14], respectively. When a photon is reflected by the wall, the $\mathbf{R}_r$ will be calculated by Lambert law, where the wall is assumed to be gray and diffuse for solar radiation.

$$\Delta e_p = e_p \cdot \beta_a / \beta_e \quad (2)$$

$$d = -(\ln \xi_1)/\beta_e \quad (3)$$

$$\mathbf{R}_r = \begin{cases} \begin{bmatrix} \cos\alpha_i \cos\gamma_i/|\sin\gamma_i| & -\cos\beta_i/|\sin\gamma_i| & \cos\alpha_i \\ \cos\beta_i \cos\gamma_i/|\sin\gamma_i| & \cos\alpha_i/|\sin\gamma_i| & \cos\beta_i \\ -|\sin\gamma_i| & 0 & \cos\gamma_i \end{bmatrix} \begin{bmatrix} \sin\theta_s \cos\varphi_s \\ \sin\theta_s \sin\varphi_s \\ \cos\theta_s \end{bmatrix}, |\cos\gamma_i| < 0.99999 \\ \begin{bmatrix} \sin\theta_s \cos\varphi_s \\ \sin\theta_s \sin\varphi_s \\ \text{SIGN}(\cos\gamma_i) \cdot \cos\theta_s \end{bmatrix}, |\cos\gamma_i| > 0.99999 \end{cases} \quad (4)$$

$$\mathbf{I}_r = \begin{bmatrix} \cos\alpha_i & \cos\beta_i & \cos\gamma_i \end{bmatrix}^T \quad (5)$$

$$\varphi_s = 2\pi\xi_2, \quad \cos\theta_s = \begin{cases} 2\xi_3 - 1 & , g = 0 \\ \dfrac{1}{2g}\left[1+g^2 - \left(\dfrac{1-g^2}{1-g+2g\xi_3}\right)^2\right] & , g \neq 0 \end{cases} \quad (6)$$

where $\mathbf{I}_r$ is the incident unit vector at the interaction site; $\xi$ is a uniformly distributed random number in the interval (0,1); $\theta_S$ and $\varphi_S$ are the deflection angle and azimuthal angle of the photon, respectively; *g* is the anisotropy coefficient, and *g*=0 when the media is isotropic; SIGN(*x*) returns 1 when *x*>0, and returns -1 when *x*<0.

Several performance parameters including the solar radiation heat source in grid *i* of the absorber $S_{sr}(i)$, solar flux absorbed in grid *j* on the wall or shined on the aperture $q_{sr}(j)$, the Local Concentration Ratio (*LCR*), the optical efficiency ($\eta_{opt}$), the reflection loss rate ($\eta_R$), the transmission loss rate ($\eta_T$) are defined in Eq.(7). The grid number used in the simulation for the absorber is 78400 (Width×height×thickness=56×56×25), and the details of the grid are shown in Fig. 4. The checked photon number is around $2\times10^{11}$ for the whole field and $1\times10^9$ for shining on the OVAR aperture.

$$S_{sr}(i) = Q_i/V_i, q_{sr}(j) = Q_j/A_j, LCR = q_{sr}(j)/DNI,$$
$$\eta_{opt} = \dfrac{Q_p + Q_w}{Q_A}, \eta_R = \dfrac{Q_R}{Q_A}, \eta_T = \dfrac{Q_T}{Q_A} \quad (7)$$

where $Q_i$ is the power absorbed in grid $i$; $Q_j$ is the power absorbed on the wall or shined on the aperture in grid $j$; $V_i$ is the volume of the grid $i$; $A_j$ is the area of the grid $j$; $Q_p$ and $Q_w$ are the power absorbed by the porous absorber and the wall, respectively; $Q_R$ and $Q_T$ are the power reflected back from the aperture and transmits through the outlet, respectively; $Q_A$ is the power shined on the aperture.

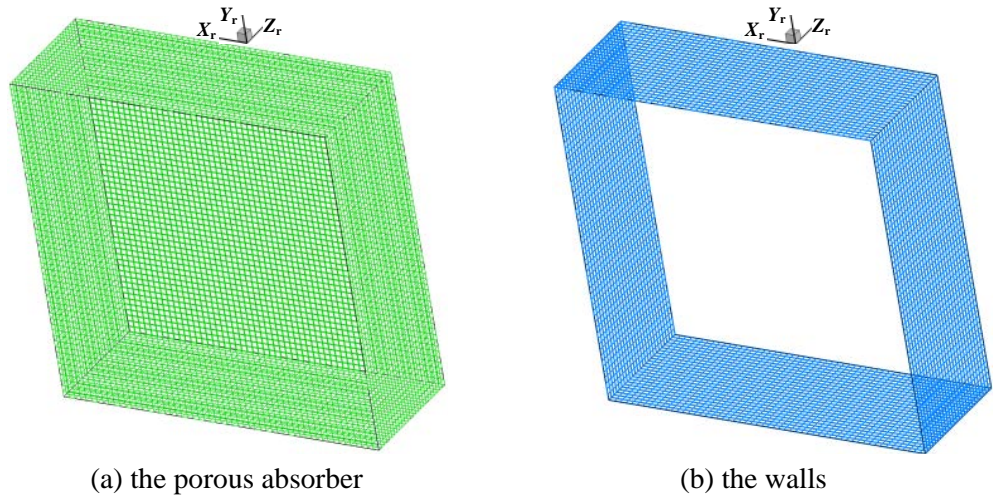

(a) the porous absorber          (b) the walls

**Fig. 4** The grids for the absorber and the walls used in the computation.

## 4. MODEL VALIDATION

To validate the model, firstly, the present *LCR* distribution computed by MCRT on the aperture plane of the OVAR is compared with that computed by SOLTRACE[1] as shown in Fig. 5, where solar azimuth $A_s=120°$, solar altitude $\alpha_s=25°$, and the atmospheric attenuation is ignored. From Fig. 5(a), it is seen that the *LCR* contours agree quite well with each other. Figure 5(b) illustrates the details of *LCR* at $X_r=0$ and $Y_r=0$, and it is found that both relative errors of the curves are within 1.3%. Then, the light transfer in a semitransparent slab was simulated, where $n=1.0$, $\beta_a=1000$ m$^{-1}$, $\beta_s=9000$ m$^{-1}$, $g=0.75$, and the slab thickness is 0.2 mm[14]. The angularly diffuse reflectance ($R_d$) and transmittance ($T_d$)[14] of the slab are shown in Fig. 6. It is observed that the present results are in good agreement with the published data.

All the above results indicate that the MCRT model is appropriate for modeling the solar radiation transfer in both the heliostat field and the porous absorber.

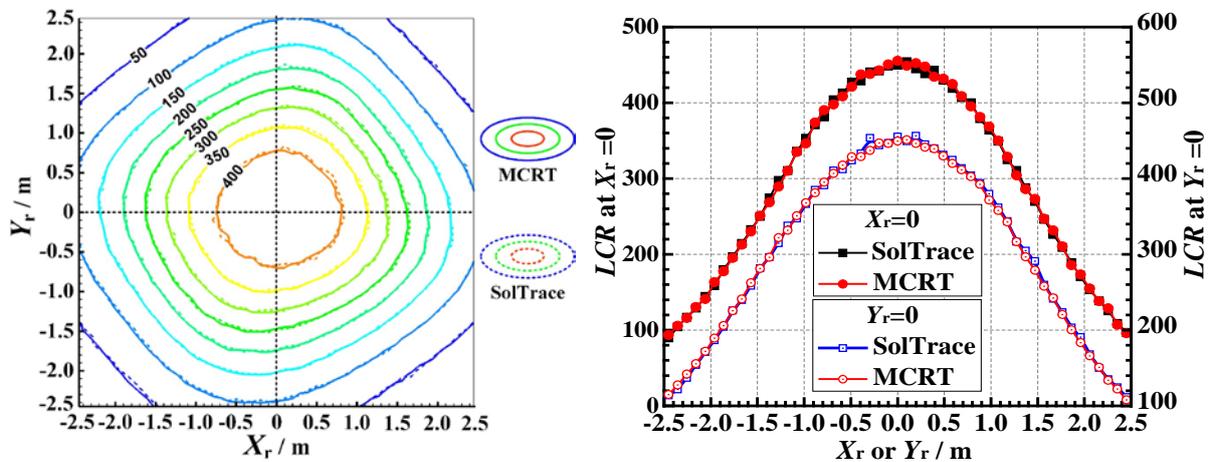

(a) *LCR* contours          (b) Details of *LCR* at $X_r=0$ and $Y_r=0$

**Fig. 5** Comparison of the *LCR* contours on the aperture plane of the OVAR between MCRT result and that of SOLTRACE.

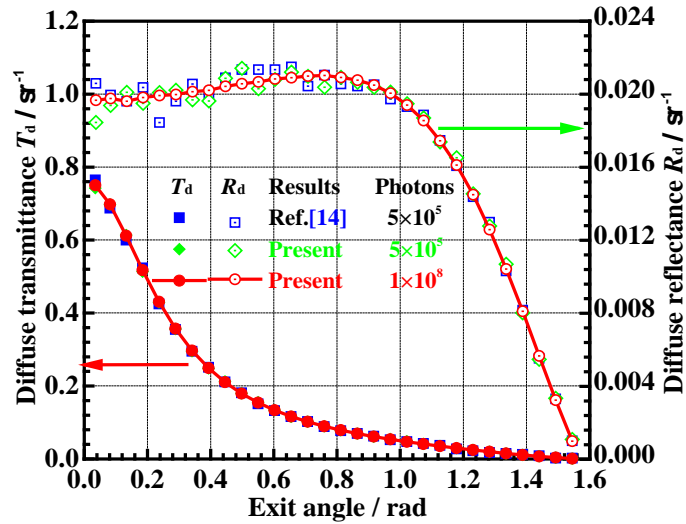

**Fig. 6** Comparisons of the angularly diffuse reflectance ($R_d$) and transmittance ($T_d$).

## 5. NUMERICAL RESULTS AND DISCUSSION

### 5.1 Solar Radiation Distribution in OVAR

Figure 7 illustrates the solar radiation distribution at the aperture plane. From Fig. 7(a), it is seen that the solar flux($q_{sr}$) on the aperture plane is non-uniform. The $q_{sr}$ decreases from the center to the margin, and the maximum flux of $2.490\times10^6$ W·m$^{-2}$ appears at the center $O$ which is the aim point of the heliostats. Figure 7(b) shows the flux at the OVAR aperture which locates at the square region in Fig. 7(a). It is seen that the highest $q_{sr}$ still appears at the center, however the largest difference of the fluxes at different locations in the aperture region is less than 1.8%. These results indicate that although the flux at the large aperture plane is significantly non-uniform, the flux at the aperture for a single OVAR is quite uniform. Figure 7(c) shows the incident angle distribution of solar rays on the aperture. It is seen that the incident angle of the non-parallel rays varies between 0° and 42°. It indicates that the parallel assumption of the rays, which is usually made in the performance analysis of OVAR, is not appropriate and should be avoided.

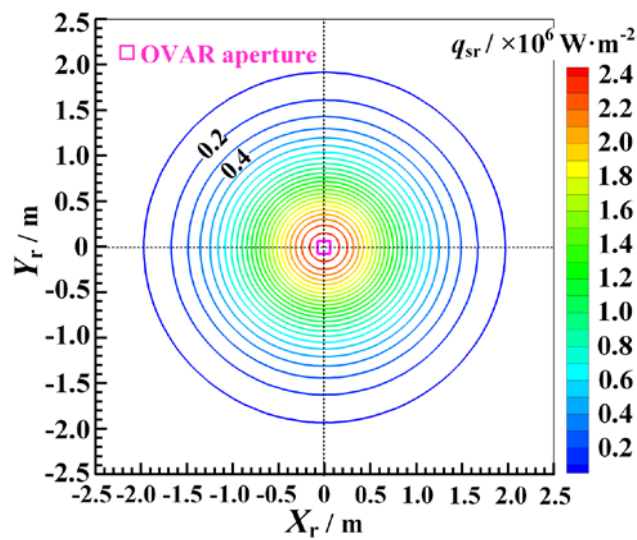

(a) Aperture plane. Incident power=$7.737\times10^8$ W

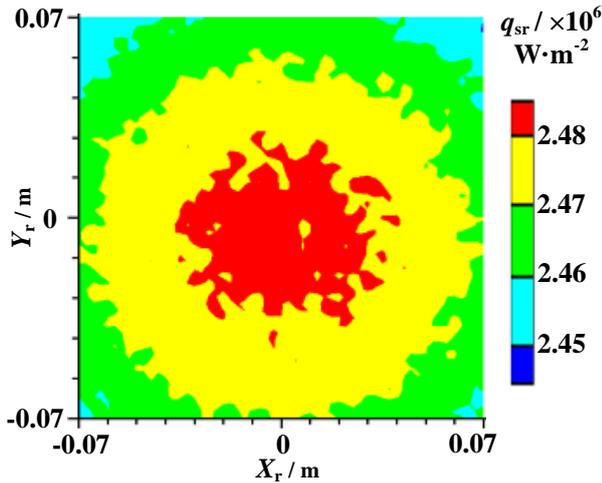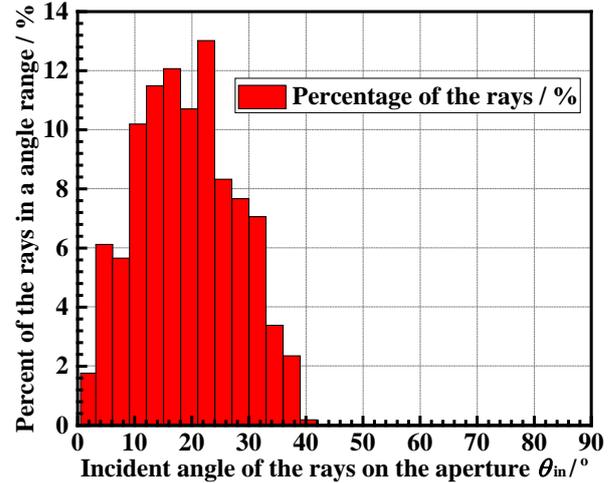

(b) OVAR Aperture.  
$q_{sr,max}$=2.490×10$^6$ W·m$^{-2}$, Incident power=48,470 W

(c) Incident angle distribution of solar rays on the aperture.

**Fig. 7.** Solar radiation distribution at the aperture of the OVAR at 12:00, spring equinox (*DNI*=961 W·m$^{-2}$).

Figure 8 illustrates the solar radiation distributions in the absorber and on the walls of the OVAR. It is seen that both the solar flux on the wall and the solar heat source ($S_{sr}$) in the porous decrease from the inlet to the outlet, because the solar radiation is absorbed gradually by the porous media along the incident direction. It is also found that the maximum $q_{sr}$ of 1.680×10$^5$ W·m$^{-2}$ and maximum $S_{sr}$ of 2.414×10$^8$ W·m$^{-3}$ appear at the inlet. The power absorbed by walls and the power absorbed in absorber are 762 W and 42,023 W, respectively.

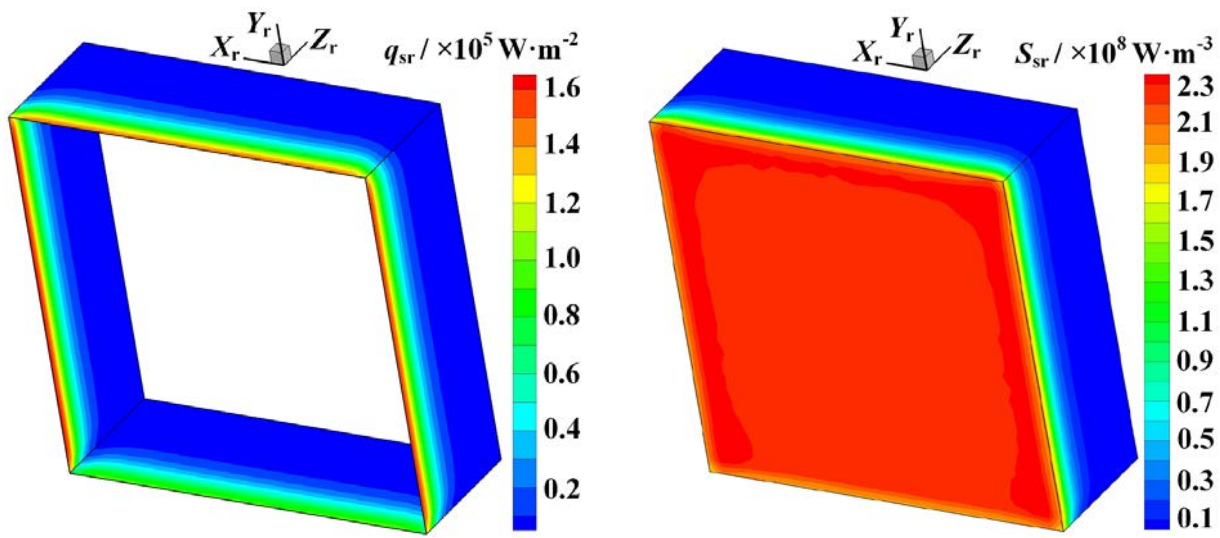

(a) Wall, absorbed power = 762 W    (b) Absorber, absorbed power = 42023 W

**Fig. 8.** Solar radiation distribution in the absorber and on the wall of the OVAR at 12:00, spring equinox (*DNI*=961 W·m$^{-2}$).

## 5.2 Effects of Non-Parallel Rays and Absorption on Wall

Figure 9 shows the effects of the non-parallel rays and the absorption on the wall on the solar heat source ($S_{sr}$) distribution in the absorber. From Fig. 9(a), it is seen that the typical $S_{sr}$ distribution at the cross section can be divided into 3 regions: (1) beside the wall region, (2) the hot spot region, and (3) the center region. It is found that the $S_{sr}$ at the beside wall region is lower than other regions, which can be explained for the reason that the incident radiation for a point at this region are mainly from the half space, e.g., point $A$ in Fig. 9 (a)), while it would be the full space, e.g., point $B$ in Fig. 9 (a) at other regions. As a result, small heat source appears beside the wall.

It is also seen that the maximum source ($S_{sr,max}$) appears at the hot spot region which is near the wall rather than the center of the receiver as usual, and this phenomenon which is counterintuitive can be explained as follows. Firstly, a point at this region can accepts radiation from the full space, so the $S_{sr}$ will not be as small as that in the beside the wall region. Secondly, the combined effects of the non-parallel incident rays and the diffuse reflection on the wall create the hot spot, because most of the rays which hit the wall will be reflected and reabsorbed by the absorber at this region. If the rays are parallel, this phenomenon will be greatly weaken as shown in Fig. 9 (b). If there is no reflection on the wall, this phenomenon will even be eliminated as shown in Fig. 9 (c). For the reason that the highest solar source locates at the region near the wall where the air velocity is lower than that in the center, so it can be inferred that the mismatching the solar source and the heat transfer ability of the air would occurs at this region, which may leads to the local overheating and results in the failure of the OVAR. The revelation of this phenomenon could offer help to the heat transfer analysis and safe operation of the OVAR in the future.

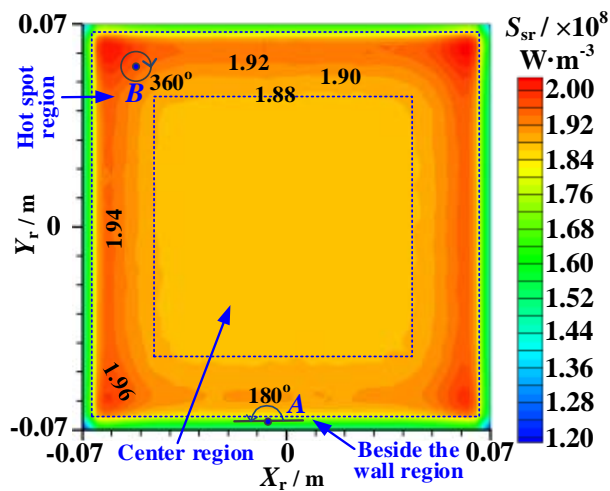

(a) Absorber under real non-parallel rays, $Z_r$ = 3mm

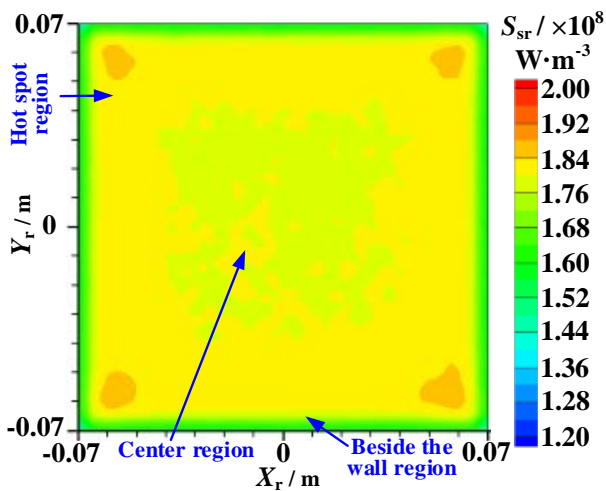
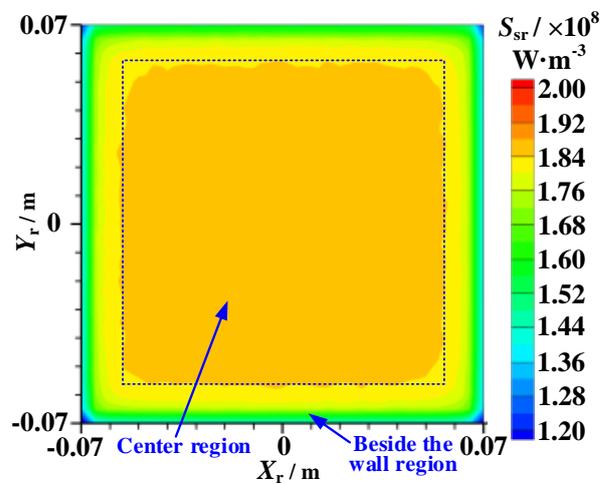

(b) Absorber under parallel-ray assumption, $Z_r$ = 3mm   (c) Absorber with wall absorptivity=1, $Z_r$ = 3mm

**Fig. 9** Effects of non-parallel rays and absorption on the wall at 12:00, spring equinox (*DNI*=961 W·m$^{-2}$).

### 5.3 Effects of Porous Absorber Parameters

The radiation properties ($\beta_a$, $\beta_s$, $\beta_e$) of the absorber depend on three parameters including the solid emissivity ($\varepsilon_p$), pore diameter ($D_p$), and the porosity ($\Phi$), and these properties determine the transfer step of the photon and the absorbed power at each site. So, for predicting the solar heat source and avoiding the local overheating which may makes the absorber burn down, the effects of the three parameters were studied. Figure 10 illustrates the effects of the three parameters on the solar heat source ($S_{sr}$) distribution at the centerline of the absorber. Table 2 shows the effects of these parameters on the receiver efficiencies.

From Fig. 10(a) and Table 2, it can be observed that the $S_{sr}$ near the inlet and the optical efficiency ($\eta_{opt}$) increase with increasing $\varepsilon_p$ due to the increasing $\beta_a$, and the $S_{sr}$ far from the inlet and the reflection loss ($\eta_R$) decrease with increasing $\varepsilon_p$ due to the decreasing $\beta_s$. It can also been seen that $S_{sr}$ decreases sharply along the centerline due to the large $\beta_e$ of the porous media, and transmission loss ($\eta_T$) is very small, i.e. $\eta_T<0.4\%$.

From Fig. 10(b), it can be seen that the maximum source ($S_{sr,max}$) decreases with increasing $D_p$. And, $S_{sr,max}$ decreases from $6.83\times10^8$ W·m$^{-3}$ for $D_p=0.5$ mm to $1.55\times10^8$ W·m$^{-3}$ for $D_p=3.0$ mm. The is because that both $\beta_a$ and $\beta_s$ decrease with increasing $D_p$, and the penetration distance of the solar photon becomes larger.

From Fig. 10 (c), it can be found that the $S_{sr,max}$ decreases with increasing $\Phi$ for the decreases in $\beta_a$ and $\beta_s$. And, $S_{sr,max}$ decreases from $8.40\times10^8$ W·m$^{-3}$ for $\Phi=0.4$ to $2.25\times10^8$ W·m$^{-3}$ for $\Phi=0.90$.

From Table 2, it can be concluded that the increases of $D_p$ and $\Phi$ influence little on the optical efficiency, respectively, owing to the simultaneous reduction in $\beta_a$ and $\beta_s$. Furthermore, the $\eta_{opt}$ of 86.70% is achieved at designed parameters, and the corresponding $\eta_R$ and $\eta_T$ are 13.20% and 0.10%, respectively.

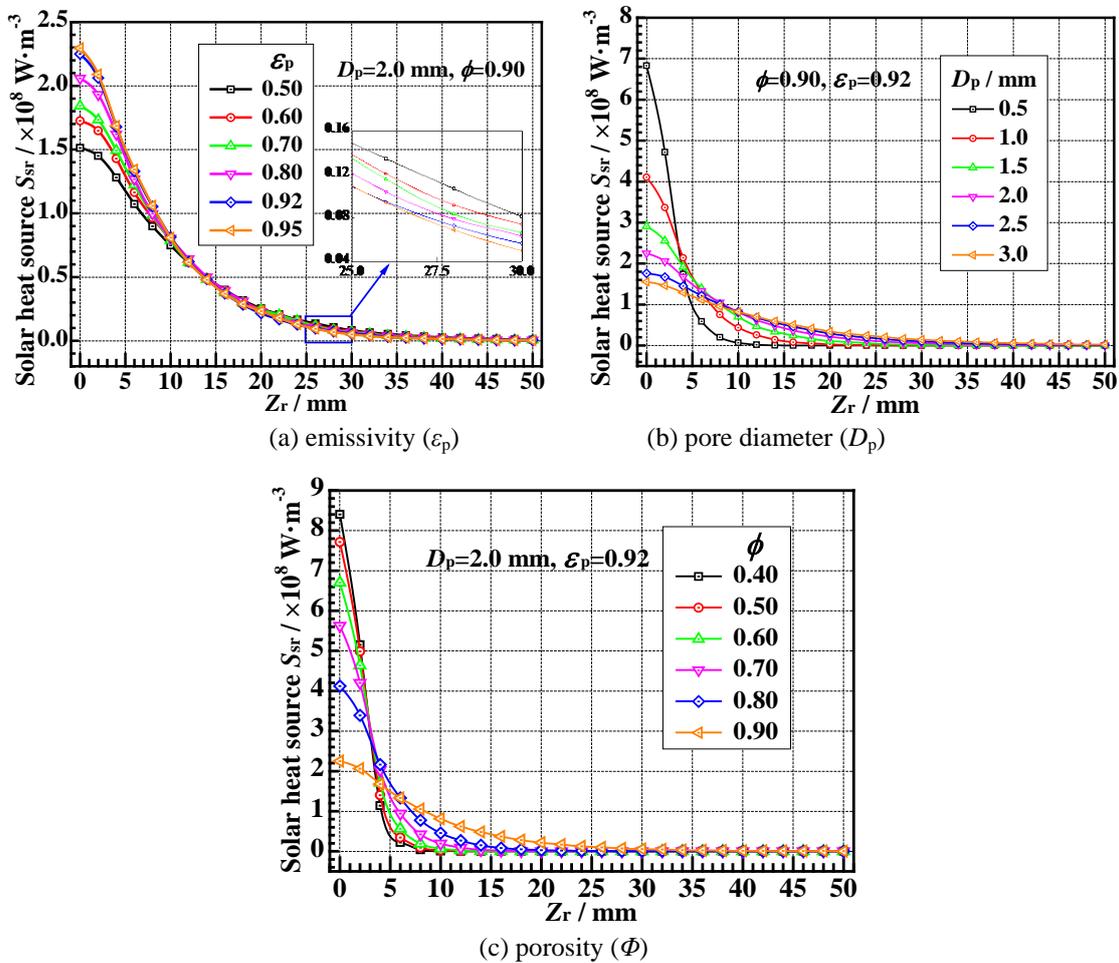

(a) emissivity ($\varepsilon_p$)  (b) pore diameter ($D_p$)

(c) porosity ($\Phi$)

**Fig. 10.** Effects of porous absorber parameters.

Table 2. Effects of solid emissivity ($\varepsilon_p$), pore diameter ($D_p$), and porosity ($\Phi$) on the receiver efficiencies.

| \multicolumn{4}{c}{$D_p$=2.0 mm, $\Phi$=0.90} | | | | \multicolumn{4}{c}{$\varepsilon_p$=0.92, $\Phi$=0.90} | \multicolumn{4}{c}{$D_p$=2.0 mm, $\varepsilon_p$=0.92} | | |
|---|---|---|---|---|---|---|---|---|---|---|
| $\varepsilon_p$ | $\eta_{opt}$/% | $\eta_R$/% | $\eta_T$/% | $D_p$ | $\eta_{opt}$/% | $\eta_R$/% | $\eta_T$/% | $\Phi$ | $\eta_{opt}$/% | $\eta_R$/% | $\eta_T$/% |
| 0.50 | 75.25 | 24.44 | 0.31 | 0.5 | 86.57 | 13.43 | 0.00 | 0.40 | 86.54 | 13.46 | 0.00 |
| 0.60 | 78.72 | 21.06 | 0.22 | 1.0 | 86.64 | 13.36 | 0.00 | 0.50 | 86.55 | 13.45 | 0.00 |
| 0.70 | 81.63 | 18.21 | 0.16 | 1.5 | 86.70 | 13.29 | 0.01 | 0.60 | 86.57 | 13.43 | 0.00 |
| 0.80 | 84.12 | 15.75 | 0.13 | **2.0** | **86.70** | **13.20** | **0.10** | 0.70 | 86.60 | 13.40 | 0.00 |
| 0.92 | 86.70 | 13.20 | 0.10 | 2.5 | 86.48 | 13.13 | 0.39 | 0.80 | 86.64 | 13.36 | 0.00 |
| **0.95** | **87.27** | **12.64** | **0.09** | 3.0 | 85.95 | 13.07 | 0.98 | **0.90** | **86.70** | **13.20** | **0.10** |

## 6. CONCLUSION

A comprehensive optical model for the SPT with an OVAR was developed and validated in this work, and the optical characteristics of the OVAR were studied based on the model. The following conclusions can be derived.

(1) The detailed distributions of the non-uniform solar flux ($q_{sr}$) on the aperture and the heat source ($S_{sr}$) in the OVAR were revealed. The flux with the maximum value of $2.490\times10^6$ W·m$^{-2}$ at the aperture is found to be relatively uniform across the aperture, however the incident angle of the rays varies between 0° and 42°, which indicates that the parallel assumption of the rays is not appropriate for the OVAR in SPT. The $S_{sr}$ in the porous absorber decreases from the inlet to the outlet, and the maximum source ($S_{sr,max}$) of $2.414\times10^8$ W·m$^{-3}$ appears at the inlet at the design parameters.

(2) The maximum source ($S_{sr,max}$) was found to appear at the region near the wall rather than the center of the receiver as usual for the combined effects of the non-parallel incident rays from the field and the diffuse reflection on the wall. The revelation of this phenomenon could offer help to the heat transfer analysis and safe operation of the OVAR in the future.

(3) Study on the effects of the porous parameters indicates that the solid emissivity influences both the $S_{sr}$ distribution and the receiver efficiency significantly. However, the pore diameter and the porosity influence the source distribution importantly, but have negligible effect on the efficiency. Moreover, the optical efficiency ($\eta_{opt}$) of 86.70%, reflection loss rate ($\eta_R$) of 13.20% and transmission loss rate ($\eta_T$) of 0.10% are achieved at designed parameters.

## ACKNOWLEDGEMENTS

The study is supported by the Key Project of National Natural Science Foundation of China(No. 51436007), the Major Program of the National Natural Science Foundation of China(No. 51590902), and the 111 Project (B16038).

## NOMENCLATURE

| | | | |
|---|---|---|---|
| $A_s$ | solar azimuth (°) | $\beta_s$ | scattering coefficient (m$^{-1}$) |
| $A_h$ | azimuth of heliostat's center normal(°) | $\beta_e$ | extinction coefficient (m$^{-1}$) |
| $DNI$ | Direct Normal Irradiance (W·m$^{-2}$) | $\varepsilon_w$ | wall emissivity ( - ) |
| $D_p$ | pore diameter (mm) | $\varepsilon_p$ | emissivity of the porous solid ( - ) |
| $d$ | travel distance of photon in porous media (m) | $\eta_{opt}$ | optical efficiency (%) |
| $e_p$ | power carried by each photon (W) | $\eta_R$ | reflection loss rate (%) |
| $G$ | tower base | $\eta_T$ | transmission loss rate (%) |
| $H$ | center of each heliostat | $\theta_h$ | heliostat azimuth in the field (rad, °) |
| $H_o$ | height of aperture center (m) | $\theta_i$ | incident angle on a surface (rad, °) |
| $\boldsymbol{I}, \boldsymbol{N}, \boldsymbol{R}$ | incident / normal / reflection vector | $\theta_S$ | deflection angle of the photon (rad) |
| LCR | local concentration ratio (-) | $\lambda_h$ | angle between the line $\boldsymbol{HO}$ and local vertical in Fig. 3 (rad, °) |
| $n$ | refractive index (-) | $\xi$ | uniformly distributed random number in the interval (0,1) |
| $O$ | aperture center | $\sigma_{te}$, $\sigma_{se}$ | standard deviation of tracking / slope error (mrad) |
| $q_{sr}$ | solar flux absorbed on wall or shined on aperture (W·m$^{-2}$) | $\varphi_S$ | azimuthal angle of the photon (rad) |
| $S_{sr}$ | solar heat source in the absorber (W·m$^{-3}$) | $\varphi$ | local latitude (°) |
| $X, Y, Z$ | Cartesian coordinates (m) | $\Phi$ | porosity ( - ) |
| $\alpha_s$ | solar altitude (rad, °) | $\omega$ | hour angle (rad, °) |
| $\alpha_h$ | altitude of heliostat's center normal (rad, °) | *Subscripts* | |
| $\alpha_r$ | altitude of the receiver (°) | g, h, r, w, i | ground / heliostat / receiver / wall / incident parameter |
| $\beta_a$ | absorption coefficient (m$^{-1}$) | p | photon / porous parameter |

# REFERENCES

[1] Garcia, P., Ferriere, A., Bezian, J.J., "Codes for solar flux calculation dedicated to central receiver system applications: A comparative review", *Sol. Energ.*, 82(3), pp. 189-197, (2008).
[2] Wang, F.Q., Tan, J.Y., Shuai, Y., Tan, H.P., Chu, S.X., "Thermal performance analyses of porous media solar receiver with different irradiative transfer models", *Int. J. Heat Mass. Tran*, 78, pp. 7-16, (2014).
[3] Roldán, M.I., Fernández-Reche, J., Ballestrín, J., "Computational fluid dynamics evaluation of the operating conditions for a volumetric receiver installed in a solar tower", *Energy*, 94, pp. 844-856, (2016).
[4] Chen, X., Xia, X.L., Meng, X.L., Dong, X.H., "Thermal performance analysis on a volumetric solar receiver with double-layer ceramic foam", *Energ. Convers. Manag.*, 97, pp. 282-289, (2015).
[5] Chen, X., Xia, X.L., Dong, X.H., Dai, G.L., "Integrated analysis on the volumetric absorption characteristics and optical performance for a porous media receiver", *Energ. Convers. Manag.*, 105, pp. 562-569, (2015).
[6] Qiu, Y., Li, M.J., He, Y.L., Tao, W.Q., "Thermal performance analysis of a parabolic trough solar collector using supercritical CO2 as heat transfer fluid under non-uniform solar flux", *Appl. Therm. Eng.*, 115, pp. 1255-1265, (2017).
[7] Rinaldi, F., Binotti, M., Giostri, A., Manzolini, G., "Comparison of linear and point focus collectors in solar power plants", *Proceedings of the Solarpaces 2013 International Conference*, 49, pp. 1491-1500, (2014).
[8] Yu, Q., Wang, Z., Xu, E., "Simulation and analysis of the central cavity receiver's performance of solar thermal power tower plant", *Sol. Energ.*, 86(1), pp. 164-174, (2012).
[9] Qiu, Y., He, Y.L., Li, P., Du, B.C., "A comprehensive model for analysis of real-time optical performance of a solar power tower with a multi-tube cavity receiver", *Appl. Energ.*, 185, Part 1, pp. 589-603, (2017).
[10] Qiu, Y., He, Y.L., Cheng, Z.D., Wang, K., "Study on optical and thermal performance of a linear Fresnel solar reflector using molten salt as HTF with MCRT and FVM methods", *Appl. Energ.*, 146, pp. 162-173, (2015).
[11] Qiu, Y., He, Y.L., Wu, M., Zheng, Z.J., "A comprehensive model for optical and thermal characterization of a linear Fresnel solar reflector with a trapezoidal cavity receiver", *Renew. Energ.*, 97, pp. 129-144, (2016).
[12] Schmitz, M., Schwarzbözl, P., Buck, R., Pitz-Paal, R., "Assessment of the potential improvement due to multiple apertures in central receiver systems with secondary concentrators", *Sol. Energ.*, 80(1), pp. 111-120, (2006).
[13] Vafai, K., Handbook of porous media, 3rd ed., CRC Press, Boca Raton, 2015.
[14] Wang, L., Jacques, S.L., Zheng, L., "MCML—Monte Carlo modeling of light transport in multi-layered tissues", *Comput. Meth. Prog. Biomed.*, 47(2), pp. 131-146, (1995).